\documentclass[twocolumn,showpacs,preprintnumbers,amsmath,amssymb]{revtex4}
\usepackage{graphicx}
\usepackage{dcolumn}
\usepackage{bm}

\begin{document}

\title{
Electronic structure, doping, order and disorder in cuprate superconductors.}

\author{T. Jarlborg}

\affiliation{
DPMC, University of Geneva, 24 Quai Ernest-Ansermet, CH-1211 Geneva 4,
Switzerland
\\}


\begin{abstract}

The electron-phonon and spin-phonon coupling in typical high-$T_C$ cuprates, 
like LSCO and HBCO are peaked for
just a few q-vectors because of the 2-dimensional Fermi surface shape.
The activation of few spin-phonon modes compensates for the low electronic
density-of-states, and the superconducting $T_C$ can be high. Thermal disorder of
the lattice perturbs the strongly coupled modes already at moderately 
high temperature. This happens because of incoherent potential fluctuations 
of the Madelung term and reduced 
spin-phonon coupling. This effect puts a limit on long-range 
superconductivity, while fluctuations can persist on a shorter length scale at higher temperatures. BCS-type
model calculations are used to show how disorder can reduce the superconducting gap
and $T_C$.  Ordering of dopants into stripes has been suggested
to improve superconducting properties, mainly through an increased DOS. Such a mechanism
seems to be a good way to make $T_C$ more resistant to thermal disorder at low doping.

\end{abstract} 

\pacs{74.20.Pq,74.72.-h,74.25.Jb}

\maketitle

\section{Introduction.}

Calculations of the electron-phonon coupling $\lambda$ (EPC) \cite{mcm} can be made 
from band parameters coming from self-consistent  electronic structure results
based on density-functional theory (DFT)
\cite{dft,gasp,daco}.
In this formalism we write the EPC 
as 
\begin{equation}
\lambda = N(E_F) I^2 / M \omega^2
\label{eqlamb}
\end{equation}
where $N(E_F)$ is the density-of-states (DOS) at the Fermi energy, $M$ is
an atomic mass, $\omega$ is the averaged phonon frequency
($M\omega^2$ is
a force constant $K = \partial^2E/\partial u^2$, the second derivative of the total energy, $E$, with
respect to the atomic displacement $\partial u^2$)
and $I$ is the matrix element $<\frac{\partial V(r)}{\partial u}>$ evaluated at the $E_F$
($V(r)$ is the electron potential).
$I$  can be determined from the change in electron
energies, $\epsilon_k$, in "frozen" phonon calculations \cite{tj6,tj7}. Such calculations show generally 
that $I$ is large only when $k$ is close to $\kappa$, the momentum for the phonon. The matrix element
is constant for harmonic vibrations, $I_k = (\epsilon_k^u - \epsilon_k^0)/u = \vartheta_k/u$.
The same formalism can be applied for coupling to spin fluctuations (SF)
essentially by replacing displacements $u$ by magnetic moments $m$ \cite{tje,tjfe}.

Most doped cuprates have very simple Fermi surfaces (FS), and retaining that
$I$ is large only for $q=\kappa$, allow us to calculate most of the coupling from a few phonon distortions only \cite{tj7}.
Furthermore, by rewriting the BCS equation for $T_C$ \cite{bcs},
\begin{equation}
k_BT_C = 1.13 \hbar\omega e^{-1/\lambda}
\label{eqbcs}
\end{equation}

 as \cite{ssc11,ssc14};
\begin{equation}
 Ku^2 = N \vartheta^2 ln (1.13\hbar \omega/ k_BT_C)
\label{eqlog2}
\end{equation}
it becomes clear that
superconductivity is possible when
the cost in total vibrational energy ($Ku^2$) of participating phonons is smaller than
a fraction (depending on the $ln$-function) of
the gain in electron energy ($N \vartheta^2$). Then, for a simple FS, it is possible to
have a large $T_C$ despite a low value of $N(E_F)$,
because the energy cost from excitations of only a few phonons will be small \cite{ssc14}. 
Phonons, spin-fluctuations and charge-waves are all possible excitations for activation of
superconductivity, but only if the electronic energy gain can overcome the excitation energy.
For conventional superconductors we usually ignore everything except phonons when $T_C$ is
to be estimated. Here, for
cuprates it is more complicated, since we assume that only part of the phonon- and spin-fluctuation
spectra should be counted.
Internal coupling within the simple FS opens the possibility for $d$-wave superconductivity
depending on the strength of electron-electron repulsion \cite{dwav}.

From such arguments we claim that the normally estimated couplings to phonons
and spin fluctuations can explain large $T_C$'s, at least quantitatively.  
Lattice deformations can influence
magnetism and other T-dependences of physical properties \cite{mj}. In the cuprates this leads to
spin-phonon coupling (SPC), where phonon distortions will enforce anti-ferro magnetic (AFM) excitations
and vice-versa 
(and $\lambda$) \cite{tj7}. Resonant inelastic X-ray scattering experiments show indications of coupling
between magnetic and lattice modes \cite{peng}. Thus, SPC might be
important for high-$T_C$ superconductivity. But the constructive SPC in the cuprates will
degrade at large T because of disorder, and if superconductivity is mediated
by few SPC modes only, it is probable that superconductivity will
suffer at high $T$ \cite{eri14}. Superconductors with complicated FS's and multiple phonon modes, 
like transition metals and their compounds,
do not have this problem, partly because SPC is not likely to be important, but also because
$T_C$ is not very high.
The simulation of EPC and SPC by frozen phonon- and spin-wave calculations 
opens the possibility how sensitive the superconductivity carrying modes are to lattice disorder. 
Here we will include thermal disorder and zero-point motion (ZPM)
in the lattice together with the phonons and spin waves in order to see how the
waves, band gaps,
coupling and $T_C$ are modified. Thermal disorder is shown to be important for ground state properties
\cite{fesi,fes1,fes2,fege,dela,js,cevib} and spectroscopic responses \cite{hed,wilk,opt,bron,dj}
in different materials. Even
ZPM is important for some properties, and since $T_C$ can be of the order 100K in cuprates we approach
a regime where band broadening becomes large in comparison to superconducting 
gaps and peaks in the density-of-state (DOS).
This will be investigated below for La$_{(2-n)}$Ba$_{n}$CuO$_4$. Finally we also present preliminary
results of how ordering of impurities into stripes can be useful to diminish the bad effects of disorder
on $T_C$ \cite{jb,apl,chm,poc,jb15}.

\section{Method of calculation.}

The band structures are calculated by the linear 
muffin-tin orbital (LMTO) method \cite{lmto,bdj} and the
local spin-density approximation (LSDA) \cite{lsda}.  
Supercells extending 2 lattice constants ($a_0$) along $\vec{y}$
 and 8 $a_0$ along $\vec{x}$ contain 112 
atoms totally, see Fig. \ref{figstr}. 
The doping $n=0.125$ (0.25 holes per Cu)
(needed to assure that waves along this cell makes
the gap at $E_F$) are taken into account in the
 virtual lattice approximation by using nuclear and electronic charges on La equal to 56.875.
The spin-polarized
 calculation include staggered fields of strengths up to $\pm 0.33$ eV on Cu sites 
 in order to induce 
 anti-ferro magnetic (AFM) stripes on the Cu lattice separated by one layer of non magnetic Cu,
 see Fig. \ref{figstr}.

 Planar oxygens are displaced by a maximum of 0.07~\AA~ in the calculations 
 with an O-"breathing" phonon mode. This displacement amplitude $u_O$ (which corresponds to thermally
 activated phonons at room temperature, RT) is large enough to obtain precise values of
 $\vartheta$ and $I$ in the harmonic approximation. The results for cells having phonon- 
 {\it and} spin-waves show a general
 enforcement of both waves, and enhanced coupling parameters, as shown previously  \cite{tj7}. In addition,
 the frozen phonon calculations show that $\vartheta_k$ is large near $k=Q-q$ 
 ($Q$ defines the AFM on near neighbor Cu sites, and $q$ defines the long-range modulation),
 and very small elsewhere.
 The calculations for ordered cells or with frozen phonons 
 will be complemented by other sets of supercells, where all atomic positions are displaced
 randomly with average displacements $u$, in order to simulate thermal disorder.

\section{Results and discussion.}
 
\subsection{Bands and coupling constants.}

\begin{figure}
\includegraphics[height=3.5cm,width=8.5cm]{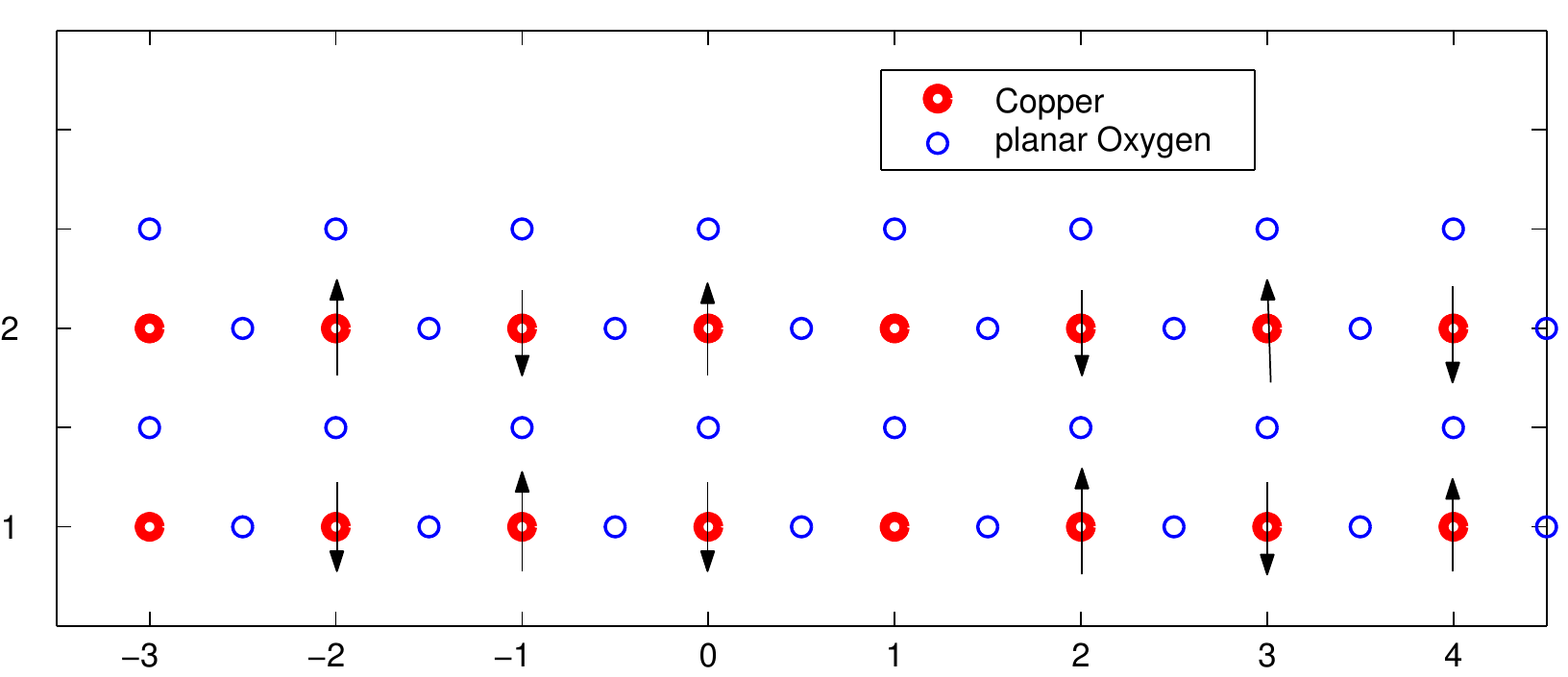}
\caption{(Color online) Schematic plot of the CuO plane
of the unit cell. The arrows indicate the applied magnetic
field on some Cu atoms that leads to AFM stripes.
}
\label{figstr}
\end{figure}

The cuprate FS's are 2-dimensional cylinders. The FS becomes
almost diamond shaped at optimal doping, when the FS reaches the X-point of the Brillouin Zone (BZ) and the DOS
has a van-Hove singularity peak \cite{eri14}. 
The calculated coupling strength for phonon/spin waves at $(Q-q)=(1/4,1/4,0)$ 
is much less efficient than for the wave in Fig. \ref{figstr}. This is because of the high local
DOS, and low band dispersion, near the $X$-point (1/2,0,0).
Therefore, the wave makes easily a gap near this point, where the coupling is large. 
The calculated $T_C$ is small. 
By including all phonons in the estimation of $\lambda = N \vartheta^2/M\omega^2u^2 \approx$ 0.06 
(from these calculations $N=0.6$ eV/Cu/spin, $\vartheta=$0.07 eV for u=0.07 \AA, and using
$M\omega^2$=10 eV/\AA$^2$ for planar oxygens \cite{tj6,tj4,tj9}) and $T_C$ is less than 1 $K$. 
The FS at optimal doping is wide near (1/2,0,0), where the Fermi velocity is low \cite{tj11},
 and becomes narrower when $k$ is approaching (1/4,1/4,0) because of larger band dispersion. The FS
and states within $E_F \pm 0.2$ eV occupy not more than
1/4th of the BZ.  
With only 1/4 of the elastic energy,
$M\omega^2u^2$, from the efficient phonons near the X-point, $T_C$ becomes of the order 15 $K$.
Higher values are obtained if spin fluctuations and SPC are included \cite{tj7}, but
all these estimations are very approximate.
It can be noted that
$T_C$ in Ba$_2$CuO$_3$ is reported to be much larger than in La$_2$CuO$_4$ \cite{liu,jin,geba}. A possible reason
could be that fewer apical oxygens make the interplane interaction weaker,
so that the FS becomes even narrower than in La$_2$CuO$_4$
\cite{jbmb,jbb}.

\begin{figure}
\includegraphics[height=6cm,width=7.0cm]{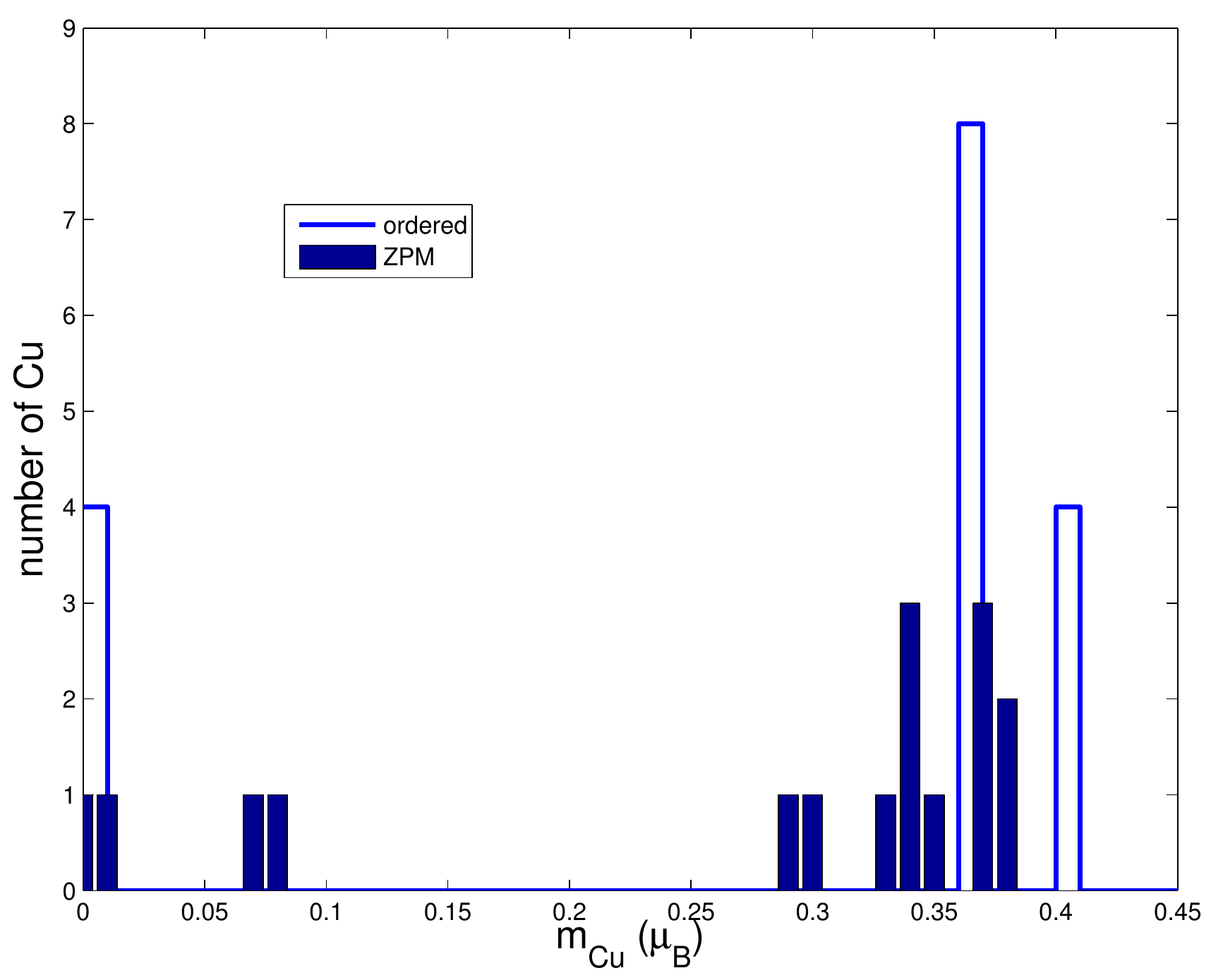}
\caption{(Color online) The distribution of the absolute values of the local Cu moments 
for ordered and ZPM-disordered structures within
supercells as in Fig. \ref{figstr}. The applied field on 12 of the 16 Cu sites is $\pm0.33$ eV in both cases.
}
\label{fighist}
\end{figure}

\subsection{Disorder and fluctuations.}

There is a negative influence from thermal disorder on the well-defined modes that convey superconductivity.
 Two sets of calculations are made with disorder, where $u$ are $0.034$ and $0.052$~\AA, 
 corresponding to ZPM and at RT
respectively. 
These distortions agree with isotropic
values on La and Cu in undoped La$_2$CuO$_4$ as measured by H\"{a}fliger et al \cite{haf}, 
while the measured values for O-sites are slightly larger.
The measured distortion of planar O in the direction of Cu (the most important for the band gap at X)
is similar to what is used here.
Our calculated broadening ($\Delta\epsilon$) of the paramagnetic bands is on the average 18 meV from ZPM 
and about 25 meV at RT, and about twice as large in spin-polarized calculations, see Table \ref{table1}. 
Photoemission by Kondo et al \cite{kon} observed broadenings for states going from 0 to 13.5$^o$ 
from the diagonal "node"
direction on the FS, to be
6-9 meV at low T and about 22 meV at 175 K. Extrapolation towards states
further away from the diagonal (and towards RT) would probably give larger values in 
better agreement with the calculated band broadening. 
For comparison we note that spectroscopic broadenings at RT in Li and Na are much
larger, 150-180 meV \cite{hed,baer}. This is because those alkali metals have very soft lattices
so that the thermal disorder is very large when $T$ is close to the melting temperature. The average
$u$ would be close to 10 percent of interatomic distances at
melting according to the Lindemann rule \cite{grim}.

A first observation is that the effective hole doping on Cu
sites diminishes with disorder. The average valence charge on Cu, 10.35 el., 
goes up by about 0.03 el. and 0.06 el. in the ZPM and RT results, respectively. These values are quite stable
to spin polarization. 
But other effects of disorder are more
important for the cuprate properties at high $T$.

\begin{table}[ht]
\caption{\label{table1}
Average magnetic moments on Cu ($m_{av}$), averaged band broadening ($\Delta\epsilon$, relative to
ordered configurations) and maximal band deviations ($\epsilon_{max}$) 
for different configurations. The latter is located at k-points near the $X$-point (and caused by phonon/spin waves),
where the DOS is highest, and can be interpreted as half of the gap value 
($\frac{1}{2}E_g$). The values of $\Delta\epsilon$ for the 3 first columns are within parentheses
since they are enhanced by the gap near the X-point. The values of $\epsilon_{max}$
for the 2 last columns are within parentheses because they are partly enhanced by disorder. Energies in meV. }
  \vskip 2mm
  \begin{center}
  \begin{tabular}{l c c c c c }
  \hline
  ~  & P-phon & AF-ord & AF+phon & AF-ZPM & AF+phon-RT   \\
  \hline \hline
 $m_{av}$    & -   & 0.192  & 0.217  & 0.186 & 0.154   \\
 $\Delta\epsilon$  & (20) & (40) & (60) & 35 & 50 \\
 $\epsilon_{max}$ & 50 & 70 & 85 & (70) & (80)  \\
  \hline
  \end{tabular}
  \end{center}
  \end{table}

\begin{figure}
\includegraphics[height=6.0cm,width=8cm]{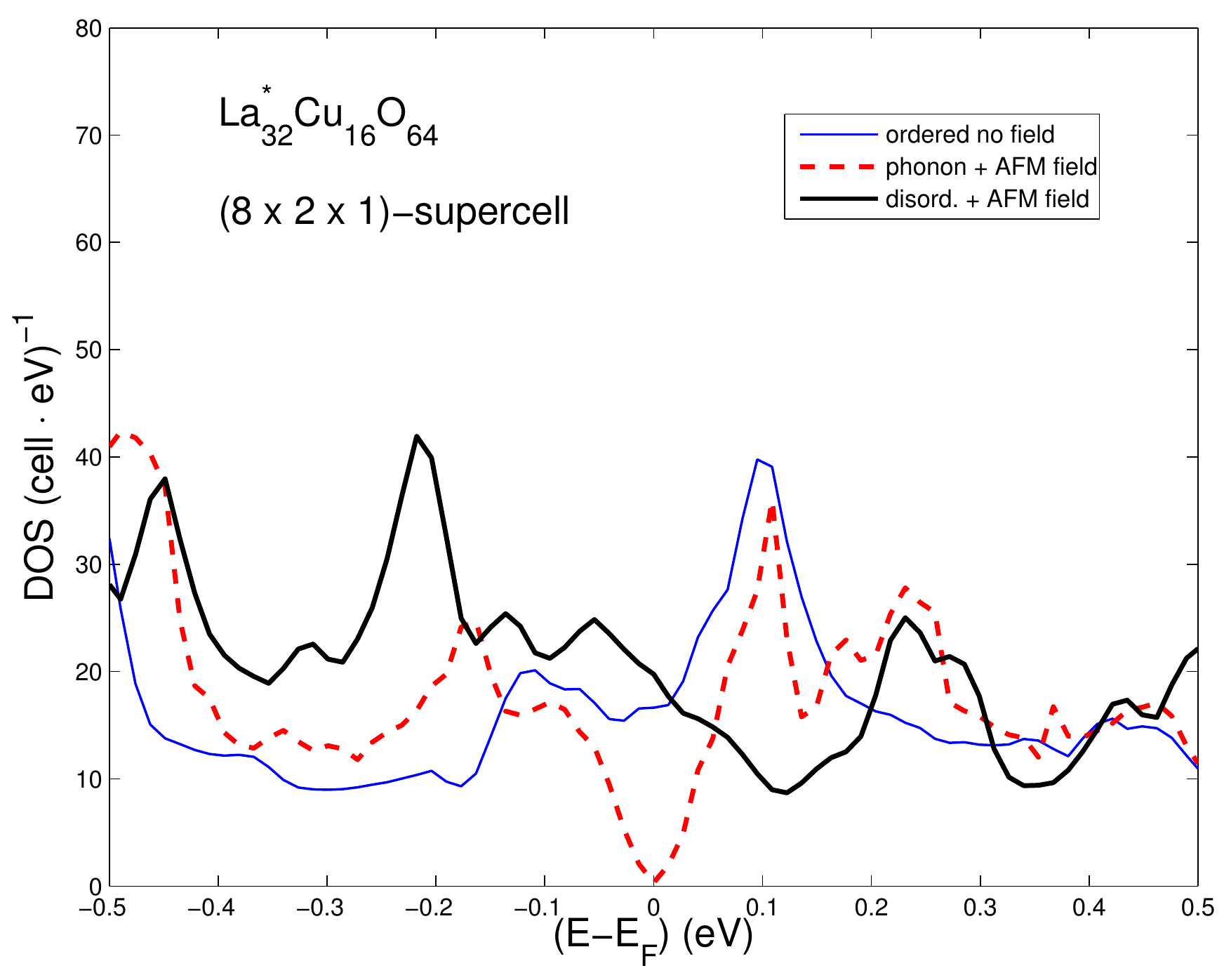}
\caption{(Color online) Unbroadened DOS functions for the non-magnetic undistorted supercell (thin blue line),
for the cell with a "breathing planar oxygen" phonon and AFM fields (bold broken red line), 
and for one disordered cell
with phonon+AFM fields (bold black line). The AFM fields induce stripes as in Fig. \ref{figstr}. The La$^*$ atoms
have charge 56.875 in these virtual crystal calculations. 
}
\label{figdos}
\end{figure}

\begin{figure}
\includegraphics[height=6.0cm,width=8cm]{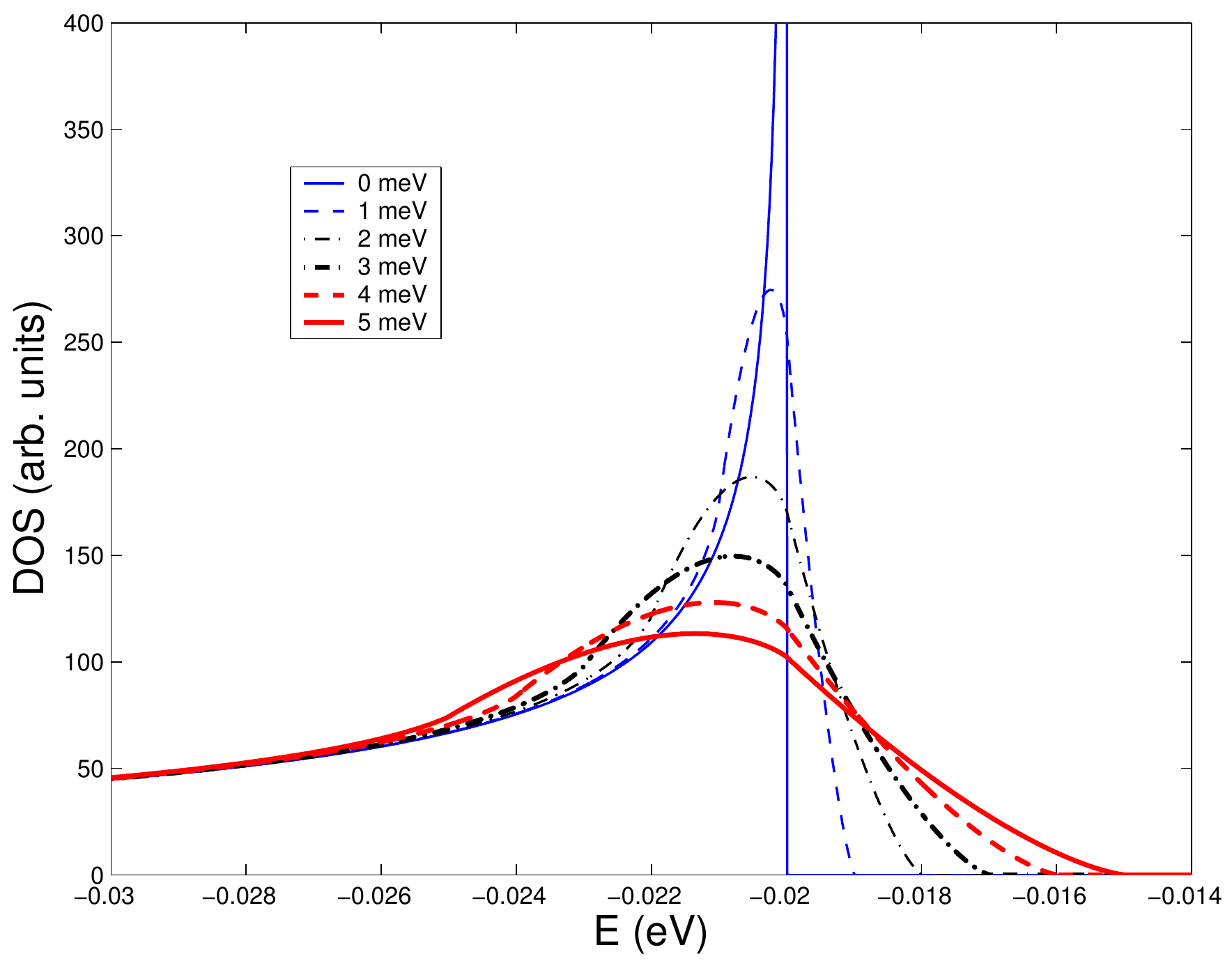}
\caption{(Color online)  Schematic low-energy edge of the BCS-like band gap as function of disorder.
}
\label{figbcs}
\end{figure}

 Fig. \ref{fighist} shows the absolute values of the local moments on Cu in the ordered and disordered 
 (at ZPM) supercells
from spin-polarized calculations with identical applied
AFM fields. The AFM wave is perfectly developed in the ordered cell with only three values of local moment amplitudes;
the four sites at node positions have no moment, the four sites at the interior
of the stripes have the largest, and the eight remaining ones have slightly smaller moments. 
This distribution of moments becomes more irregular for increasing disorder. At the AFM nodes (Cu without
field) small moments develop with disorder. It is seen that the average
moment decreases with disorder for long-wave spin modulations and for large moments despite the
positive contributions of $\frac{dm}{dT}$ that always come from the node sites. This is also shown in Table \ref{table1}, even when
both spin waves and phonons are present.  

Differently, results for shorter waves (in cells extending 4$a_0$ and 6$a_0$ along $\vec{x}$) show that disorder
can enforce weak local moments. This is similar to what is found for disordered Cerium \cite{cevib}
compared to ordered fcc Ce \cite{ce}, but only at the volumes near the onset of magnetism. A local
volume expansion can increase the local moment, whereas a local compression cannot decrease an already
weak moment, and on the average there is an increase \cite{cevib}. Overdoped cuprates are weakly magnetic
and they have positive values of $\frac{dm}{dT}$, at least for $m$ up to 0.4 $\mu_B$/Cu. A positive
average of $\frac{dm}{dT}$ can also be found in weak moment configurations with long spin waves due to strong 
contributions from node sites.  Not however that disorder  destroys 
coherent waves in all cells, and the gaps at $E_F$ are more and more difficult to detect
at ZPM and RT in the cells of different lengths and different average moments.

The effect of SPC, demonstrated by the higher moment
in a lattice with a phonon distortion compared to the moment in the ordered structure, 
is weaker in the disordered lattices. This is rather independent of doping.

The coupling constant for SF appears to be more sensitive to disorder than
the coupling for phonons, because the "clean" AFM wave suffers a lot 
in the disordered lattice. 
Instead
of having a potential perturbation given by one Fourier coefficient at a precise $q$, there will
be several coefficients, each one working to open up gaps at different parts of the band.  
This has a direct negative impact on the gap.
The maximal band deviation, $\epsilon_{max}$,
is found at k-points near the position of the downfolded $X$-point, i.e. in the region of the BZ where
the single phonon/spin wave opens up a gap. The two disordered configurations have 
large band broadenings at all k-points. On the contrary, at k-points where a well-defined wave
makes a gap, disorder tends to wipe out the gap. In fact, $\epsilon_{max}$ is not revealing for the
latter cases, since it might be boosted from disorder just at one k-point.

  The closing of the gap in the disordered lattice is seen from
 the DOS functions in Fig \ref{figdos}. The thin line
  shows the DOS for the non-distorted non-magnetic supercell. The $N(E_F)$ is quite high since $E_F$
  is close to the van-Hove singularity for this hole doping, although the DOS calculated 
from the limited number of k-points is not smooth. With the phonon and the spin-wave
  along the cell there is a gap and $N(E_F)=0$, as shown by the broken line.
 This DOS curve changes to the bold line when all atomic positions are disordered, and there is
  no sign of a gap any longer. Peaks and valleys in the DOS appear at other energies for this particular
  (randomized) disorder, but the superimposed DOS from several different disordered configurations
 becomes more smooth.

 The next step is to project the parameters of band broadening from the
ab-initio results on to the BCS DOS function at different $T$ in order to extract the effect of
disorder on the DOS peak, on the superconducting gap ($\Delta$) and on $T_C$. This is exemplified in Fig. \ref{figbcs}.
The DOS peak below $E_F$ becomes broadened, but its energy is essentially unchanged. 
 The upper band edge
will reach the lower band edge above the gap when $\Delta$ decreases at higher $T$. Superconductivity
disappears approximately when $\Delta-\Delta\epsilon$=0, although  
it could survive longer through a mechanism of percolation. 
Small disconnected
islands of fluctuating superconductivity can appear temporarily, and still produce a DOS-peak below a 
pseudogap.  The gap defined from the peak in scanning
tunneling microscopy (STM)  
can be larger than 50 meV in underdoped cuprates with the highest $T_C$'s,
and the gap ratio ($2\Delta/k_BT_C$) is typically much larger than the BCS-value (3.5) even in
cuprates with lower $T_C$ \cite{stm}.
Suppose that $\Delta\approx$40 meV at low $T$, as often found from STM for cuprates \cite{stm}. 
 This corresponds to a $T_C$ near $250K$ according to BCS theory.
The ZPM makes 
the DOS-peak broad. In the spin-polarized results for the supercell
we have a broadening of about 30 meV from ZPM, which makes a reduction of the band edge to $\sim 10$ meV. 
At 150K, when a normal BCS gap would be reduced to $30-35$ meV,  
 the broadening increases to about 40 meV. The gap would be closing at
 $\sim 120K \approx 11 meV$, which would be the corrected $T_C$ in this example. 
Therefore, thermal disorder will reduce $T_C$ more
than gap (defined from the DOS peak at 0K), which leads to a larger gap-ratio than in BCS. In the example
above it would be 7.5, and the DOS-peak remains visible for $T > T_C$.
Obviously, these results are only qualitative, but the general conclusions
are; $^1$ that thermal disorder can produce superconducting fluctuations and a pseudogap above $T_C$, 
$^2$ that the gap-ratio is larger than the BCS value in underdoped cuprates,
and $^3$ that a higher $T_C$ is expected
if disorder could be suppressed.

Band calculations show that the coupling parameter for spin-fluctuations
 goes up for decreased doping (more than for phonons), and it tends to
diverge before static AFM makes the material insulating \cite{tj6}.
Therefore, it is on the underdoped side of the phase diagram that we can expect the strongest
disorder effects on superconductivity.
 There is a competition between large possible $T_C$
and its destruction from disorder, because both mechanisms depend
on the coupling parameter for spin fluctuations. The absence of superconductivity at low $T$ and very low doping
can be because the increased coupling makes disorder important already from ZPM, while this
is not the case at higher doping. 
This makes sense since the effect of disorder is accumulated over a long distance and therefore
more destructive for long waves at low doping.
Another possibility is that
 the increasingly long wave-lengths of the modulations at low doping
make superconductivity sensitive to the short coherence length \cite{eri14}. 
Long-range superconductivity is replaced by fluctuations in both cases.
 
 \subsection{Making $T_C$ resistent to disorder?}

The simple relations of how $\lambda$ and $T_C$ depend on $N$, $\vartheta$ and $Ku^2$, indicate
ways for increasing $T_C$ if disorder plays no role.   
A different view of the high-$T_C$ problem, and how to increase $T_C$, is possible if it turns out
that the pseudogap at low doping is caused by superconducting fluctuations in a thermally disordered lattice.
One could then recuperate a higher $T_C$
through stabilization of existing superconducting fluctuations, but 
it is difficult to get rid of thermal disorder. Larger coupling
constants for spin fluctuations should be efficient to enforce the "clean" spin waves. 
However, a large $\vartheta_{m}$ would also enforce fluctuations
and bad influence from
disorder could spoil the effect of enforced spin waves. Structural disorder will be smaller in a stiffer lattice
at a given $T$, and since the spin disorder is largely an effect of the structural disorder, it suggests that
lattice hardening would help. This can be achieved by applying pressure, $P$, as long as the strength of the spin wave
 is preserved.  A higher $N(E_F)$ leads to a 
larger gain in electronic energy from a superconducting gap, and  
$T_C$ would be higher according to eq. \ref{eqlog2}. This is also valid at large disorder.
Broadening of DOS peaks, which are not exactly at $E_F$ at low $T$, can be favorable for
increasing $N(E_F)$ at large $T$. Another way to 
increase $N$ is to introduce
new (static) potential modulations in the lattice \cite{jb,poc,chm}. 
Such  modulations could be caused by (natural or artificial)
dopant orderings, which make peaks in the DOS at $E_F$ if the period is correct \cite{apl}.
In this case we imagine that the FS is not going to be more complicated, but that the band at $E_F$
becomes more flat.
Even efforts to {\it {decrease}} $\vartheta_m$ 
in order to avoid static AFM, or suppress AFM fluctuations, in the 
very underdoped region could have a positive effect on $T_C$. This can be achieved by applied pressure.
In fact, $dT_C/dP$ appears to be positive more so at low doping than in the overdoped region \cite{schill},
even if other $P$-dependent variations (phase transitions, charge carrier transfers within the structure)
can also modify $T_C$.

\section{Conclusion.}

The conclusion from these results is that the effective $\lambda's$ can larger than what is 
generally believed. This is because the superconducting mechanism depend on few
phonon/spin excitations. However, gaps and long-range phonon/spin waves are sensitive to disorder and
there is a limitation of long-range superconductivity and $T_C$ because of thermal disorder. 
This is most likely at low doping where
the coupling from spin-fluctuations is largest. A softened lattice may increase $\lambda$, but it also 
makes the amplitude of disorder larger. To increase $N(E_F)$ by letting $E_F$ to be on a DOS-peak should
increase the effective $\lambda$ even if $\vartheta$ is constant. This could therefore be a way to reach
a higher $T_C$ through a moderation of the harmful effects from thermal disorder. 

{\bf Acknowledgement} I am grateful to A. Bianconi and C. Berthod for useful discussions.

\end{document}